\documentclass[aps,superscriptaddress,amsmath,amssymb,reprint,longbibliography]{revtex4-1}
\usepackage{color}
\usepackage{epsfig}
\usepackage{hyperref}
\usepackage{graphicx}
\usepackage{dcolumn}
\usepackage{bm}
\usepackage[dvipsnames]{xcolor}
\usepackage[normalem]{ulem}

\begin{document}

\title{Detection and amplification of spin noise using scattered laser light in a quantum-dot microcavity}

\author{A.~N.~Kamenskii}
\affiliation{Experimentelle Physik 2, Technische Universit\"at Dortmund, 44221 Dortmund, Germany}

\author{M.~Yu.~Petrov}
\affiliation{Spin Optics Laboratory, Saint Petersburg State University, 198504 St.\,Petersburg, Russia}

\author{G.~G.~Kozlov}
\affiliation{Spin Optics Laboratory, Saint Petersburg State University, 198504 St.\,Petersburg, Russia}

\author{V.~S.~Zapasskii}
\affiliation{Spin Optics Laboratory, Saint Petersburg State University, 198504 St.\,Petersburg, Russia}

\author{S.~E.~Scholz}
\affiliation{Angewandte Festkörperphysik, Ruhr-Universit\"at Bochum, 44780 Bochum, Germany}

\author{C.~Sgroi}
\affiliation{Angewandte Festkörperphysik, Ruhr-Universit\"at Bochum, 44780 Bochum, Germany}

\author{A.~Ludwig}
\affiliation{Angewandte Festkörperphysik, Ruhr-Universit\"at Bochum, 44780 Bochum, Germany}

\author{A.~D.~Wieck}
\affiliation{Angewandte Festkörperphysik, Ruhr-Universit\"at Bochum, 44780 Bochum, Germany}

\author{M.~Bayer}
\affiliation{Experimentelle Physik 2, Technische Universit\"at Dortmund, 44221 Dortmund, Germany}
\affiliation{Ioffe Institute, Russian Academy of Sciences, 194021 St.\,Petersburg, Russia}

\author{A.~Greilich}
\affiliation{Experimentelle Physik 2, Technische Universit\"at Dortmund, 44221 Dortmund, Germany}

\begin{abstract}
Fundamental properties of the spin-noise signal formation in a quantum-dot microcavity are studied by measuring the angular characteristics of the scattered light intensity. A distributed Bragg reflector microcavity was used to enhance the light-matter interaction with an ensemble of $n$-doped (In,Ga)As/GaAs quantum dots, which allowed us to study subtle effects of the noise-signal formation. Detecting the scattered light outside of the aperture of the transmitted light, we measured the basic electron spin properties, like $g$-factor and spin dephasing time. Further, we investigated the influence of the microcavity on the scattering distribution and possibilities of signal amplification by additional resonant excitation.
\end{abstract}

\maketitle

In recent years optical spin noise spectroscopy (SNS) has developed into an efficient research tool in the field of spin physics~\cite{Mueller10, VSZRev}.
Initially demonstrated in thermal vapors of alkali atoms~\cite{Alexandrov81, Crooker04}, it has been further applied to spins in bulk and low-dimensional semiconductor structures~\cite{Oestreich05, Mueller08, Crooker09, CrookerPRL2010, DahbashiPRL14}, and recently extended to studies of the valley dynamics in monolayer semiconductors~\cite{Goryca19} and the magnetization fluctuations in ultrathin metal films~\cite{CrookerPRX18}.

In optical SNS, the fluctuations of the magnetization close to the ground state are mapped onto Faraday rotation angle fluctuations using magneto-optical effects~\cite{GiriPRB12}. In other terms, the spin noise signal arises from an interference of the forward-scattered field with the transmitted driving laser~\cite{Gorbovitskii83,GlazovOE15,ScalbertPRB19}. Therefore,  understanding of the scattering gives a direct link to the properties of the studied system~\cite{STSNSPRL19}.

In general, the measured spin noise signal is proportional to the probe beam intensity squared.
Thus, increasing intensity, on the one hand, improves the sensitivity of the measurements but, on the other hand, increases unwanted perturbations of the system~\cite{DahbashiAPL12,GlasenappPRB13}. One possibility to decrease these perturbations could be the use of optical resonators, in particular, microcavities, which can be considered as an efficient tool of signal amplification~\cite{Zapasskii2011,PoltavtsevPRB14,DahbashiPRL14}. This possibility is, however, not optimal, as the increased light-matter interaction leads to an increased perturbation of the system, so that a strong reduction of the light intensity is required. Furthermore, additional limitations are given by the diode detectors, which limit the sensitivity of the recorded signal at low level optical intensities by their own  electrical noise. A solution could be provided by the basic properties of the spin noise signal formation, i.e. by the fact that a coherent superposition of the driving laser field with the forward-scattered field is equivalent to implementing a homodyne detection. In this case, the laser transmitted through the sample can be replaced by a part of the laser beam, which is not going through the sample and therefore does not interact with the system. This allows one to use a very low probe power for accessing the spin noise while working with high power laser light hitting the diodes~\cite{Cronenberger16,Sterin18,Homodyne2018}.

In this paper, we examine the abilities of homodyne detection and study the spatial properties of the scattered light.
To test the spatial distribution of spin noise we vary the angle of incidence of the probe laser on the $n$-doped quantum dot (QD) ensemble while interfering (homodyning) the scattered light along the direction of normal incidence on the sample with a reference beam (local oscillator, LO). We then apply a two-beam geometry to study the possibility of amplification of the scattered light intensity~\cite{KozlovPRA17}.

\begin{figure}[t]
\includegraphics[clip,width=\columnwidth]{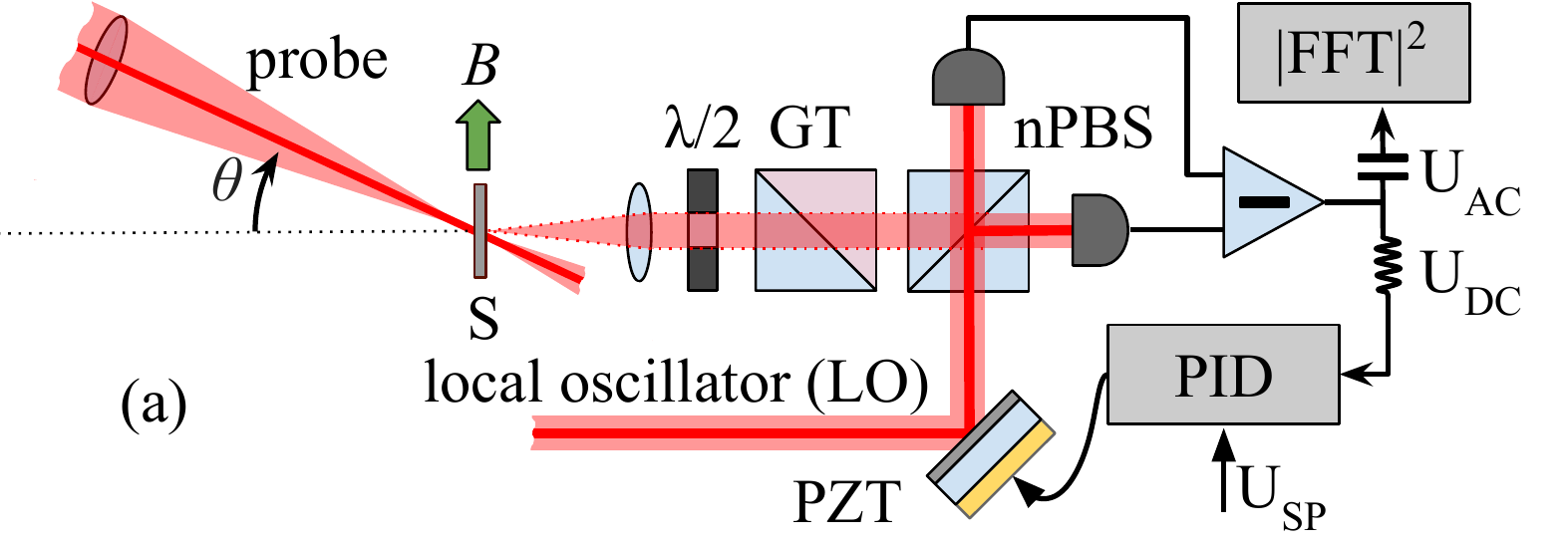}
\includegraphics[clip,width=\columnwidth]{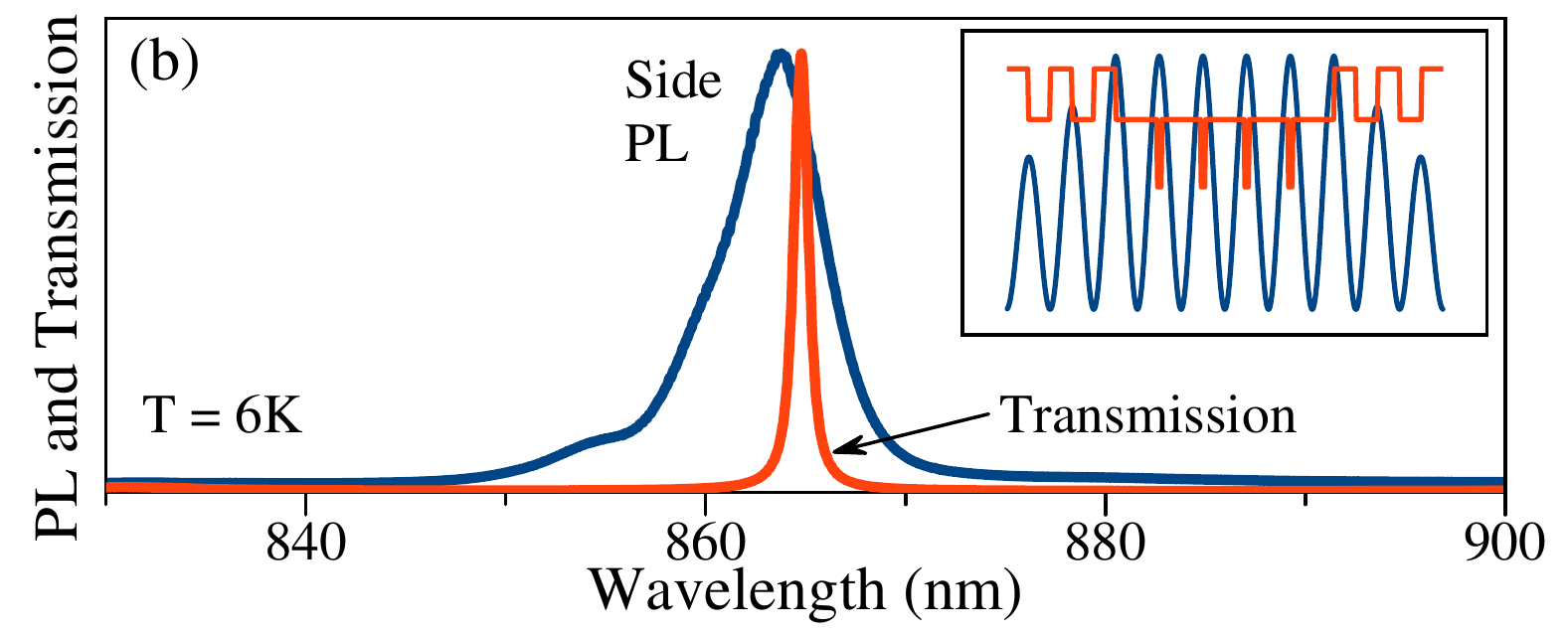}
\caption{(a) Scheme of homodyne detection setup. The probe beam hits the sample (S) under the angle $\theta$.
The interference of the electric fields of passed scattered light and local oscillator (LO) is measured at the photodiodes of the balanced receiver. The current difference is converted into the voltage signal $U(t)$. The \textit{ac}-component of $U(t)$ is used to collect the spin noise power spectrum, while the \textit{dc}-component is used for the piezo control in PID-loop. The relative optical phase is maintained by tuning the piezo-actuated mirror to the setpoint $U_\mathrm{SP}$. (b) QD PL measured along the cavity plane for 785\,nm excitation (blue line) and cavity transmission (red line) measured for white light excitation. Inset is sketch of the $5\lambda/2$ cavity showing the electric field distribution (blue line) and $4$ QD layers placed at the field antinodes seen as reduced potential levels. The red line gives the potential of the DBR structure with QDs.}
\label{fig:One}
\end{figure}

Figure~\ref{fig:One}(a) shows the scheme of the spin noise experiment using homodyne detection.
The emission of a single-mode frequency-stabilized laser is split into two beams by a polarizing beam splitter representing the input of a Mach-Zehnder interferometer (not shown)~\cite{Homodyne2018}. The vertically linear polarized probe beam hits the sample (S) under the angle $\theta$ relative to incidence normal. We use focusing and collimating lenses of $200$\,mm (40\,$\mu$m spot diameter) and $60$\,mm focal length, respectively.
As the transmitted and scattered light have orthogonal linear polarizations, we use a half-wave plate ($\lambda/2$) and a Glan-Taylor prism (GT) to filter out the transmitted light component~\cite{GlazovOE15}. A fraction of the scattered light, namely the fraction within the solid angle covered by the numerical aperture $\mathrm{NA} = 0.07$ of the collimating lens, is selected from the whole $4\pi$ distribution, and directed to the input of a 50:50 non-polarizing beam splitter (nPBS), together with the laser sent through second arm (LO) of the interferometer. The interference of the two fields, having both linear horizontal polarizations, results in the photocurrents in the balanced photoreceiver ($670$~MHz bandwidth), where their difference current is converted into the voltage signal $U(t)$. The \textit{ac}-component of $U(t)$ is digitized using a $2$~GS/s analog-to-digital converter and Fourier transformed using an FPGA-based real-time fast Fourier algorithm processing~\cite{CrookerPRL2010}. The \textit{dc}-component is sent to the error input of the  proportional integral derivative (PID) control loop used to adjust the piezo voltage~\cite{Homodyne2018}. Thereby, the relative optical phase between the two arms of the interferometer can be maintained by tuning the piezo-actuated mirror (PZT) to the set point $U_\mathrm{SP}$.

As discussed in Ref.~\cite{KozlovPRA17}, spin diffusion of carriers reduces the signal for higher wave vector values or, in our configuration, for larger values of $\theta$. Therefore, to eliminate this contribution, we designed a strongly localized electron system: we use an ensemble of $n$-doped (In,Ga)As/GaAs QDs, grown by molecular-beam epitaxy. The QDs are embedded in a distributed Bragg reflector (DBR) structure with a quality factor $Q \approx 10^3$, in order to enhance the Faraday rotation and scattering intensity. A moderate $Q$-factor value was specially designed to avoid non-linear effects present in the high-$Q$ cavities~\cite{RyzhovJAP15,Scalbert2016}.
The structure was annealed at $900^{\circ}$ Celsius to shift the ground state emission of the QDs to the energy of the cavity transmission.
The $5\lambda/2$ cavity ($\lambda$ is the design wavelength) with $14$ bottom and $11$ top pairs of AlAs/GaAs stacks is optimized for the transmission geometry and contains four QD layers each with a density of $10^{10}$~cm$^{-2}$, positioned at the antinodes of the electric field at a distance of $129$\,nm between the layers, see the inset of Fig.~\ref{fig:One}(b). To provide $n$-doping for the QDs, we placed layers of Si-dopants $64.5$~nm below each dot layer. Figure~\ref{fig:One}(b) demonstrates the cavity transmission (red line) and the QD photoluminescence (PL) spectrum (blue line), detected along the cavity plane (not filtered by the cavity) for above barrier excitation at $785$~nm. Both lines overlap close to the PL maximum. The optical frequency of the probe is tuned to the microcavity resonance at $864.69$~nm. The sample is mounted on the cold finger of a helium-flow cryostat where it is cooled down to $5$~K. An external magnetic field $B$ can be applied by an electromagnet orthogonal to the cavity, as shown in Fig.~\ref{fig:One}(a).

\begin{figure}[t]
\includegraphics[clip,width=\columnwidth]{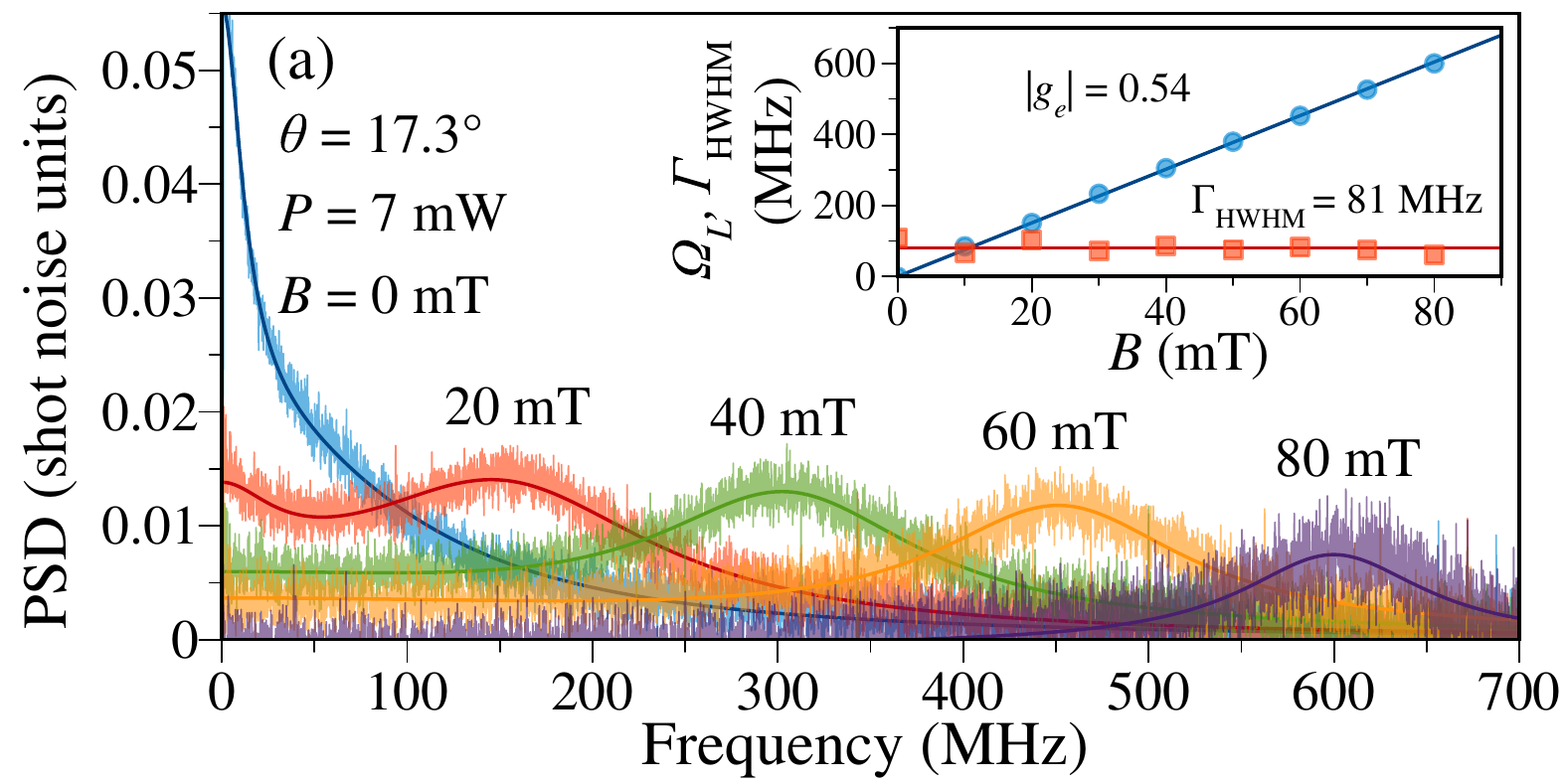}
\includegraphics[clip,width=\columnwidth]{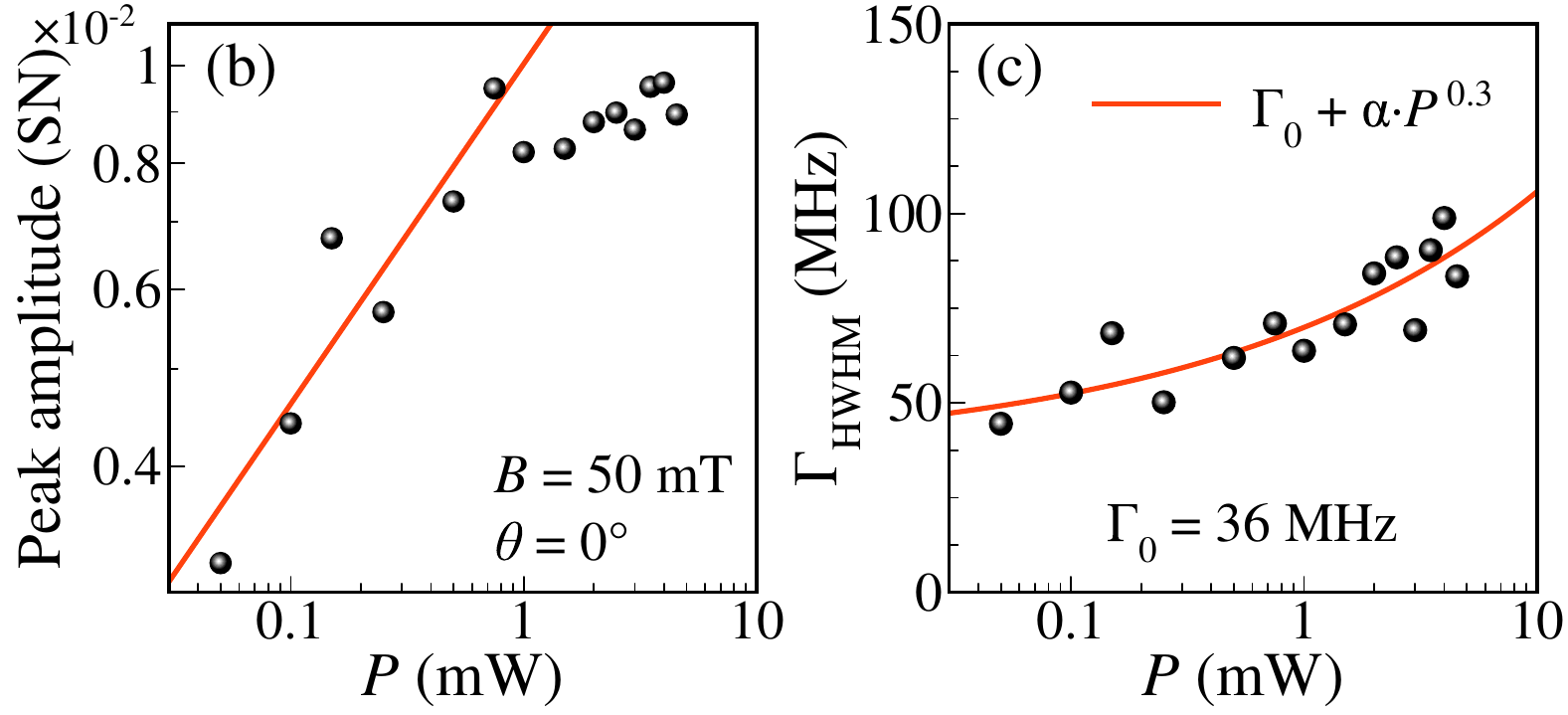}
\caption{(a) Magnetic field dependence of the spin noise for $\theta=17.3^{\circ}$. Probe power $P=7$~mW and $P_\text{LO}=4$~mW. Inset demonstrates $B$-dependence of the peak position B with a linear fit, corresponding to the electron $g$~factor $|g_e|=0.54$, and of the half-width of the peak, which is constant over the measured range, $\Gamma_\text{HWHM} = 81$~MHz. (b) Power dependence of spin noise amplitude of the Larmor peak at $\theta=0^{\circ}$ for $B=50$~mT in a log-log plot, showing also a linear fit. (c) Peak width as function of probe power with a power law fit yielding the minimal width $\Gamma_0=36$~MHz at zero-power cutoff.}
\label{fig:Two}
\end{figure}

Figure~\ref{fig:Two}(a) demonstrates an exemplary measurement of the spin noise for different magnetic fields at a probe incidence angle of $\theta = 17.3^\circ$. After transmission through the sample, the probe with wavelength $\lambda = 864.69$~nm does not enter the $\lambda/2$ and the GT-prism in the detection arm. The light scattered by the carriers along the direction normal to the sample is directed to the nPBS, see Fig.~\ref{fig:One}(a).
This result uniquely shows the observation of spin noise outside the aperture of the transmitted probe, confirming the theoretical results of Refs.~\cite{Gorbovitskii83,KozlovPRA17}, and supports a recent report in Ref.~[\onlinecite{STSNSPRL19}].

The noise signal consists of a double Lorentzian-peak structure, with a peak centered around zero frequency and a second peak, moving proportionally to the magnetic field (Larmor peak)~\cite{CrookerPRL2010, GlasenappPRB16}.
Here, we only concentrate on the magnetic-field dependent Larmor peak, which allows us to extract basic parameters of the system.
The peak position vs magnetic field gives the average $g$-factor $|g_e|=0.54$, similar to the QDs measured in Ref.~\cite{CrookerPRL2010}, blue points in inset of Fig.~\ref{fig:Two}(a). The half width $\Gamma_\text{HWHM}$ of the Larmor peak at half-maximum (HWHM) defines the spin lifetime $T_s = 1/(2\pi\Gamma_\text{HWHM}$), where $1/T_s = 1/\tau_s + 1/\tau$. $\tau_s$ is the spin relaxation time and $\tau=n_0/G$ is the recombination time, which depends on the carrier concentration, $n_0$, and the generation rate of carriers, $G$~\cite{DzhioevPRB02,SmirnovPRB18}.
In the range of magnetic fields where the peak position appears at frequencies below $1$~GHz, the Larmor peak width stays constant at $\Gamma_\text{HWHM} = 81$~MHz corresponding to $T_s = 1.96$~ns, as shown by the red points in the inset of Fig.~\ref{fig:Two}(a).

To extract the electron spin relaxation time $\tau_s$, which is not affected by the probe excitation, we measure the power dependence of the Larmor peak, see Figs.~\ref{fig:Two}(b) and~\ref{fig:Two}(c), where the dependence is shown for $\theta=0^{\circ}$, $\lambda = 864.69$~nm, and $B = 50$~mT.
Figure~\ref{fig:Two}(b) gives a log-log plot with the power dependence of the peak amplitude demonstrating that the signal saturates at powers above $1$~mW. Using the power law fit to the data in Fig.~\ref{fig:Two}(c), we extrapolate $\Gamma_\text{HWHM}$ to zero power from which we obtain the intrinsic width $\Gamma_0 = 36$\,MHz, which corresponds to $\tau_s=4.4$~ns. Further, it is known that $\tau_s$ is given by the spread of $g$-factors $\Delta g$ and the fluctuating nuclear fields in the electron surrounding, $\Delta B_N$~\cite{CrookerPRL2010,LiPRL12,GlasenappPRB16} as $\Gamma_0 = (2\pi\tau_s)^{-1} = (2\pi\hbar)^{-1}\sqrt{(\Delta g \mu_B B)^2+(g \mu_B \Delta B_N)^2}$. Here, the $\mu_B$ is the Bohr magneton and $\hbar$ is the reduced Planck constant. As the peak width is constant in the measured range of fields, we can set $\Delta g = 0$, and determine the $\Delta B_N=\hbar/(\tau_s g\mu_B)=4.8$~mT, in accord with the value taken from QDs without DBR structure, see Ref.~[\onlinecite{GlasenappPRB16}].
Note that to achieve a weakly perturbative regime of probing, the probe power for measuring the spin noise of the  QDs in the DBR structure is reduced by two orders of magnitude as compared to that in a bare QD ensemble. In this case, homodyne detection is the only way to measure the spin-noise signal while working with shot-noise limited photodetection~\cite{Sterin18,Homodyne2018}.

\begin{figure}[t]
\includegraphics[clip,width=\columnwidth]{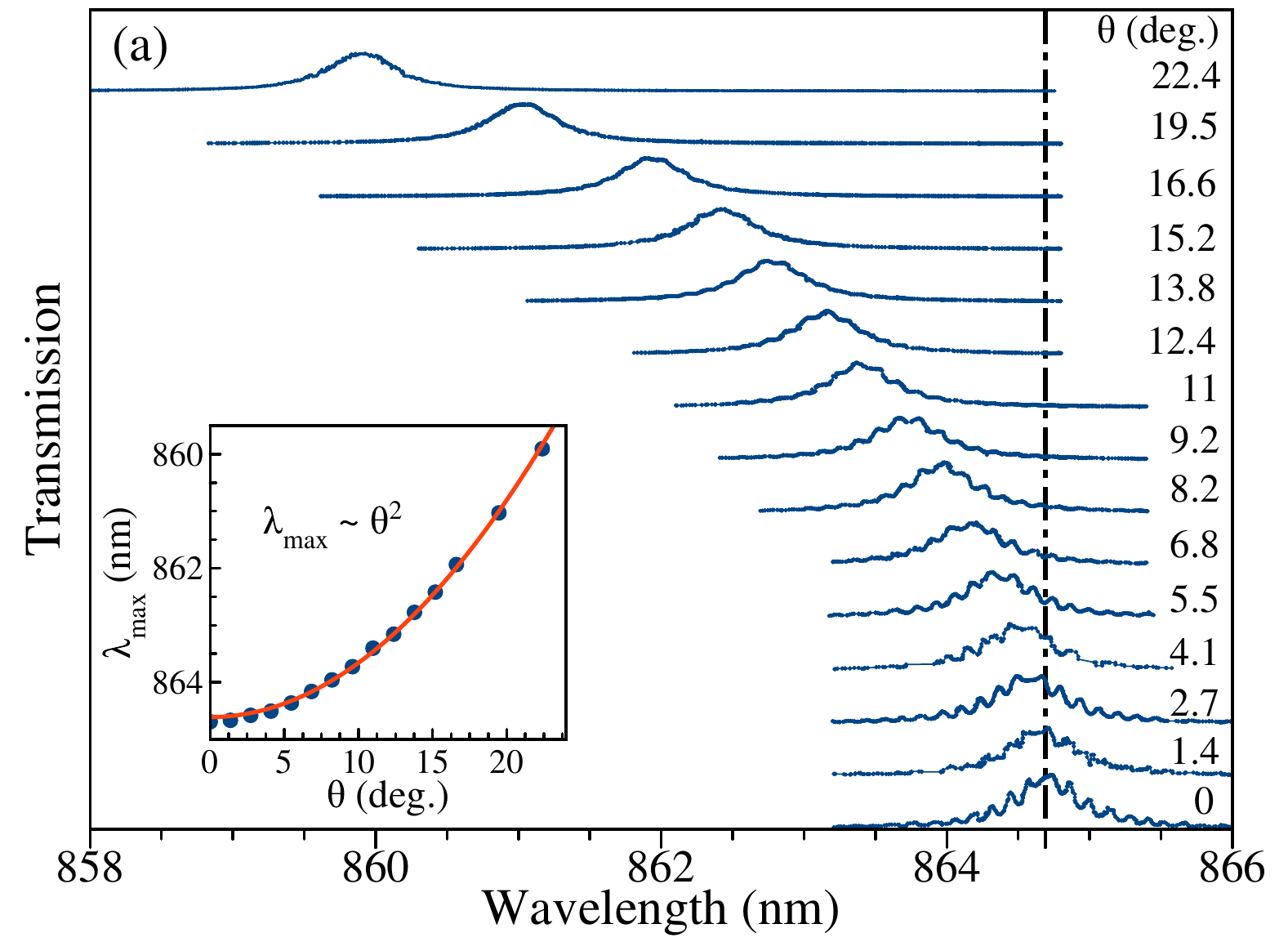}
\includegraphics[clip,width=\columnwidth]{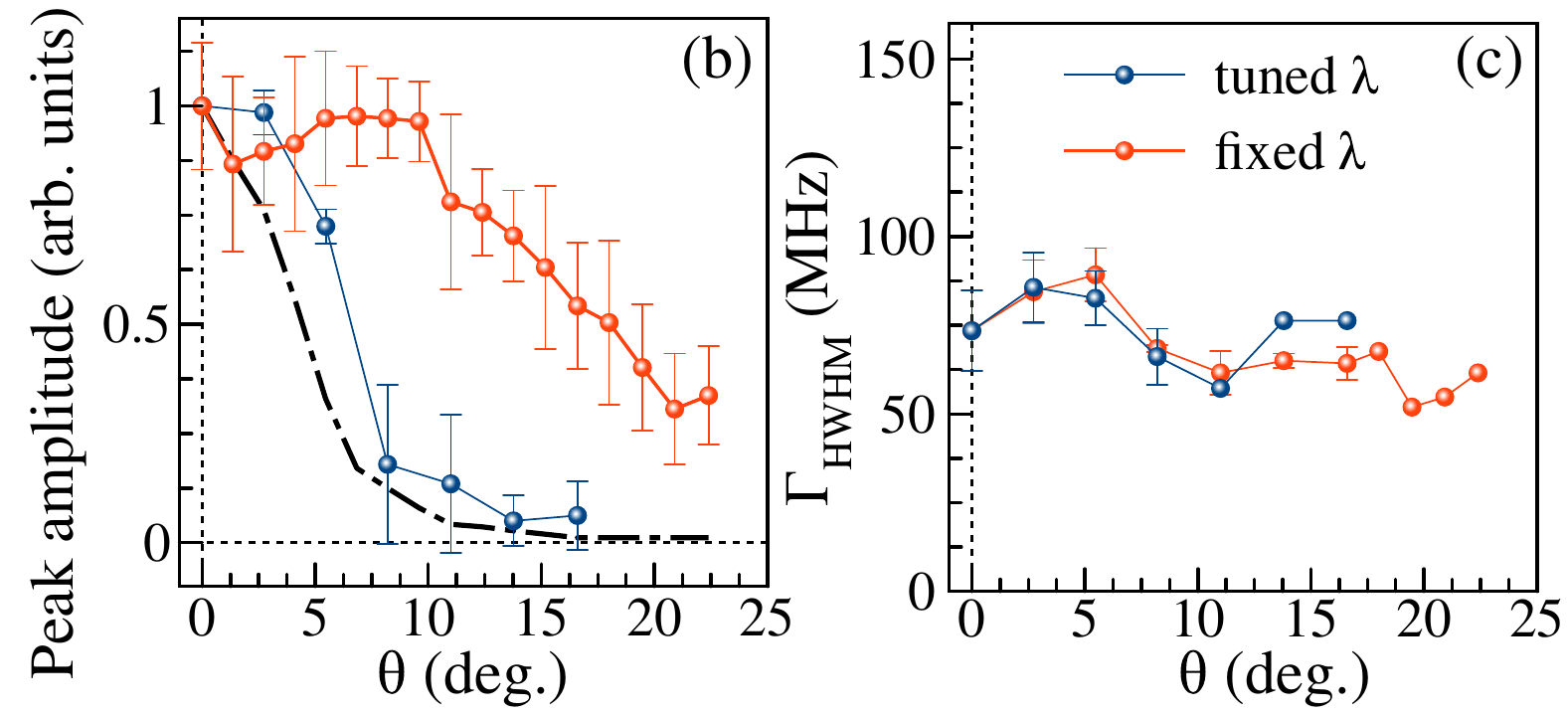}
\caption{(a) Angle dependent laser transmission of the DBR microcavity. The black dash-dotted line marks the position of the transmission maximum $\lambda_\text{max}=864.69$~nm at $\theta = 0^{\circ}$. Inset shows transmission peak position ($\lambda_\text{max}$) versus angle $\theta$ fitted by parabolic function, red line. (b)~Normalized amplitude of the spin-noise Larmor peak as function of probe incidence angle $\theta$ for two cases: red points for probe wavelength fixed at $\lambda_\text{max}(\theta = 0^{\circ}) = 864.69$~nm, blue points for probe wavelength tuned to $\lambda_\text{max}$ of each angle, see inset in panel (a). Black dash-dotted line corresponds to transmission intensity at fixed $\lambda = 864.69$~nm, as marked in the panel (a). (c) Corresponding peak widths as function of $\theta$. Lines are guides to the eye. Error bars are calculated by averaging over three independent measurements. $P = 2.3$~mW and $P_\text{LO} = 4$~mW. $B = 50$~mT.}
\label{fig:Three}
\end{figure}

In a next step, we analyze the effect of the DBR microcavity on the transmitted laser intensity as function of the angle of incidence $\theta$.
For this purpose, we first test the cavity transmission by modulating the probe laser by a mechanical chopper in front of the sample and detecting the transmitted intensity using a single silicon diode and a lock-in amplifier. By scanning the laser, the intensity of the transmitted light is recorded as function of wavelength for different angles $\theta$, see Fig.~\ref{fig:Three}(a). The inset in the figure demonstrates the expected parabolic angle dependence of the energy of maximal transmission. Figures~\ref{fig:Three}(b) and~\ref{fig:Three}(c) demonstrate the angular dependences of the measured spin noise amplitude and $\Gamma_\text{HWHM}$ of the Larmor peak at $B=50$~mT, respectively.
Here, we have considered two cases: (i) the probe wavelength is fixed at $\lambda = 864.69$~nm, corresponding to the case of maximal transmission at $\theta = 0^{\circ}$ (red points), and (ii) the probe wavelength is shifted to the corresponding transmission maximum for each angle (blue points), as given in the inset of panel (a).

Let us recall the angular dependence of the transmitted light shown in Fig.~\ref{fig:Three}(a).
If the probe laser wavelength is fixed at maximal transmission for $\theta=0^{\circ}$, then by increasing the angle the laser intensity reaching the QDs is continuously decreasing, as shown by the dash-dotted line in Figs.~\ref{fig:Three}(a) and~\ref{fig:Three}(b). However, the emission of the scattered light for this wavelength is most efficient in the direction normal to the sample (direction of homodyne light detection).
Our estimations show that for our DBR structure the scattered light is collected within a solid angle of $\sim 8^\circ$ (NA$=0.07$), defined by the microcavity $Q$~factor, see Fig.~\ref{fig:Three}(a). Therefore, the behavior in the first case can be understood by the laser power reaching the QDs through the DBR. At angles $\theta < 10^{\circ}$, the spin noise amplitude does not change a lot, as the QD excitation is still efficient and close to the saturation power. In this range, the transmitted power decreases from $100$\% to about $5$\%, or from $2.3$~mW to about $0.1$~mW, see Fig.~\ref{fig:Two}(b). For higher angles, the excitation continues to drop to levels, at which it is strongly suppressed so that the noise peak amplitude drops. The peak width also decreases proportionally to the reduction of excitation power, as seen in Fig.~\ref{fig:Three}(c).

The second case is different, see the blue points in Figs.~\ref{fig:Three}(b) and~\ref{fig:Three}(c).
Here, the excitation is always efficient, up to $100$\% of the light is reaching the QDs.
However, the extraction of the scattered light along the normal direction becomes less efficient and follows the intensity profile given by the dash-dotted line in Fig.~\ref{fig:Three}(b). In other words, with adjustment of the wavelength to the cavity transmission maximum, the scattered light exits the microcavity under the same angle as the laser does. This means that its fraction along the normal direction, where the collimation is performed, decreases. We also conclude at this stage that a cavity of high $Q$~factor strongly limits the scattering aperture for detection at a fixed wavelength, which reduces the range of possible wave vectors and hinders experiments like the one proposed in Ref.~\cite{STSNSPRL19}.

\begin{figure}[t]
\includegraphics[width=\columnwidth]{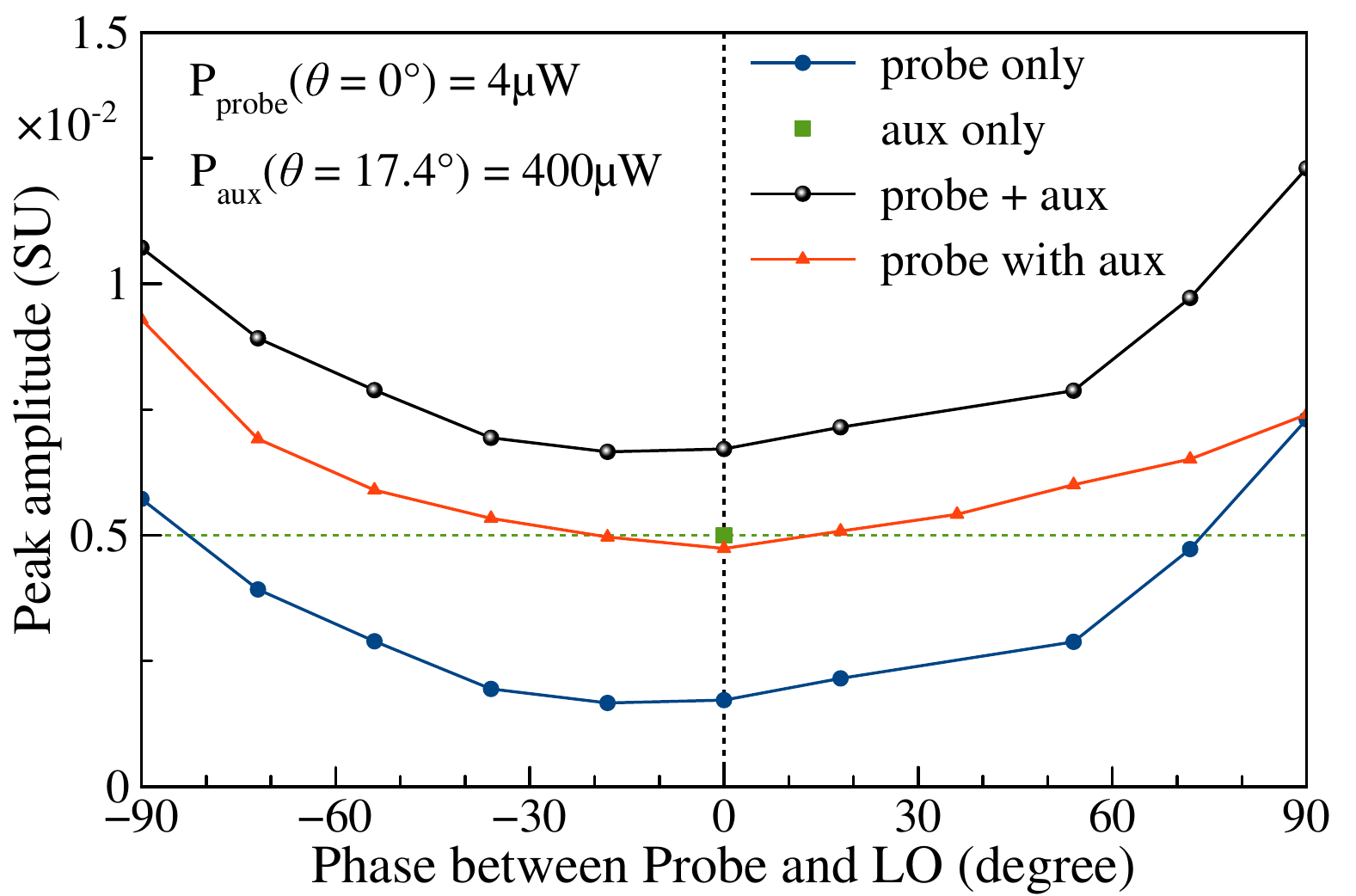}
\caption{Spin noise peak amplitude as function of phase with (red points) and without (blue points) additional excitation by an auxiliary beam. The green square at zero phase represents a measurement of the Aux only beam without phase stabilization (probe beam is closed). Black circles give the addition of the signals of Probe and Aux, measured separately. Lines are guides to the eye. $\lambda = 864.69$~nm for all beams, $P_\text{LO}=4$~mW, and $B = 50$~mT.}
\label{fig:Four}
\end{figure}

An additional effect may influence the amplitude of the spin noise in the second case. For larger angles, the transmission wavelength is shifting to shorter values, which leads to excitation of different QD sub-ensembles: as one can extract from Figs.~\ref{fig:One}(b) and~\ref{fig:Three}(a), an angle variation from $0^{\circ}$ up to $10^{\circ}$ shifts the wavelength down by about $1$~nm. According to the PL spectrum, the cavity shifts from the low energy side of the PL maximum to the high energy one, where the spin noise amplitude usually does not change a lot~\cite{GlasenappPRB16}.
In the experiment, on the contrary, the spin noise amplitude vanishes at angles of about $\theta = 10^{\circ}$, which makes this effect not relevant in our case.

Furthermore, we tested the possibility to amplify the scattered light intensity by additional resonant excitation with a second laser beam, as discussed in Ref.~\cite{KozlovPRA17}. For this purpose we applied the probe laser under $\theta = 0^{\circ}$ and an additional auxiliary (aux) beam at $\theta = 17.4^{\circ}$, which was split from the same laser and had the same polarization as the probe. Figure~\ref{fig:Four} demonstrates the results of these measurements. The blue points give the amplitude of the spin noise peak at $B = 50$~mT as function of the phase difference between probe and LO, without the aux beam~\cite{Homodyne2018}. The green point at zero phase is result of a measurement of the aux beam only.
Here, we did not stabilize the phase between the LO and the aux beam, and the probe beam was blocked. If we add up the two separately measured signals, we obtain the data points shown by the black circles. This situation corresponds to the best case for both beams applied together~\cite{KozlovPRA17}. However, our measurement results in the data points presented by the red triangles, demonstrating a non-trivial dependence of the auxiliary amplification on the relative phase between probe and LO. At $90^{\circ}$ phase difference the auxiliary beam has no effect at all on the scattered light (red and blue points overlap), while at $-90^{\circ}$ the effect is almost completely additive. This shows that the additional auxiliary excitation can increase the scattering but has its own phase dependence relative to the probe and LO beams. As discussed in Ref.~\cite{KarraiSM03}, the scattered photons consist of two components, the coherent part and incoherent part. The coherently scattered photons have the same spectral properties as the excitation laser, and can therefore interfere with the driving laser or the LO. The observed phase dependence between the scattered photons from the probe and aux beams allows us to conclude that the photons, originating from the same laser source, scatter coherently at the QDs. Additionally, to be able to see any effect of the scattering amplification, the powers of the probe and aux beams should be in a range where the QD transitions are not saturated by the laser excitation. If one of the beams has higher power, it dominates the effect and no noticeable amplification can be observed. This observation demonstrates that in the low power case the photons are scattered in the Heitler regime with weak perturbation of the system~\cite{NguyenAPL11,MatthiesenPRL12,ScalbertPRB19}.

To summarize, we have shown that the spin noise signal is determined by the properties of the scattered light, which is strongly modified by the DBR microcavity. This effect has to be carefully considered, especially for high-$Q$ microcavities used for polariton condensates. We have found that the spin noise amplitude can be amplified by additional illumination, while the system is not saturated by any excitation. The relative optical phase between the probe and auxiliary beams has to be taken into account for understanding the spin noise, which supports that the technique is observed in the non-perturbative regime.

We are grateful to S.~V.~Poltavtsev for valuable advices regarding the cavity design. We acknowledge financial support by the Deutsche Forschungsgemeinschaft in the frame of the International Collaborative Research Center TRR 160 (Project A5) and the Russian Foundation for Basic Research (Grant No. 19-52-12054). The Dortmund team acknowledges support by the BMBF-project Q.Link.X (Contract No. 16KIS0857). The Bochum team thanks the support by the BMBF-project Q.Link.X (Contract No. 16KIS0867), and the DFH/UFA CDFA-05-06. Support by the Saint-Petersburg State University through Grant No. 40847559 is also acknowledged.


\begin{thebibliography}{33}%
\makeatletter
\providecommand \@ifxundefined [1]{%
 \@ifx{#1\undefined}
}%
\providecommand \@ifnum [1]{%
 \ifnum #1\expandafter \@firstoftwo
 \else \expandafter \@secondoftwo
 \fi
}%
\providecommand \@ifx [1]{%
 \ifx #1\expandafter \@firstoftwo
 \else \expandafter \@secondoftwo
 \fi
}%
\providecommand \natexlab [1]{#1}%
\providecommand \enquote  [1]{``#1''}%
\providecommand \bibnamefont  [1]{#1}%
\providecommand \bibfnamefont [1]{#1}%
\providecommand \citenamefont [1]{#1}%
\providecommand \href@noop [0]{\@secondoftwo}%
\providecommand \href [0]{\begingroup \@sanitize@url \@href}%
\providecommand \@href[1]{\@@startlink{#1}\@@href}%
\providecommand \@@href[1]{\endgroup#1\@@endlink}%
\providecommand \@sanitize@url [0]{\catcode `\\12\catcode `\$12\catcode
  `\&12\catcode `\#12\catcode `\^12\catcode `\_12\catcode `\%12\relax}%
\providecommand \@@startlink[1]{}%
\providecommand \@@endlink[0]{}%
\providecommand \url  [0]{\begingroup\@sanitize@url \@url }%
\providecommand \@url [1]{\endgroup\@href {#1}{\urlprefix }}%
\providecommand \urlprefix  [0]{URL }%
\providecommand \Eprint [0]{\href }%
\providecommand \doibase [0]{http://dx.doi.org/}%
\providecommand \selectlanguage [0]{\@gobble}%
\providecommand \bibinfo  [0]{\@secondoftwo}%
\providecommand \bibfield  [0]{\@secondoftwo}%
\providecommand \translation [1]{[#1]}%
\providecommand \BibitemOpen [0]{}%
\providecommand \bibitemStop [0]{}%
\providecommand \bibitemNoStop [0]{.\EOS\space}%
\providecommand \EOS [0]{\spacefactor3000\relax}%
\providecommand \BibitemShut  [1]{\csname bibitem#1\endcsname}%
\let\auto@bib@innerbib\@empty
\bibitem [{\citenamefont {M\"uller}\ \emph {et~al.}(2010)\citenamefont
  {M\"uller}, \citenamefont {R\"omer}, \citenamefont {H\"ubner},\ and\
  \citenamefont {Oestreich}}]{Mueller10}%
  \BibitemOpen
  \bibfield  {author} {\bibinfo {author} {\bibfnamefont {G.~M.}\ \bibnamefont
  {M\"uller}}, \bibinfo {author} {\bibfnamefont {M.}~\bibnamefont {R\"omer}},
  \bibinfo {author} {\bibfnamefont {J.}~\bibnamefont {H\"ubner}}, \ and\
  \bibinfo {author} {\bibfnamefont {M.}~\bibnamefont {Oestreich}},\ }\href
  {\doibase 10.1103/PhysRevB.81.121202} {\bibfield  {journal} {\bibinfo
  {journal} {Phys. Rev. B}\ }\textbf {\bibinfo {volume} {81}},\ \bibinfo
  {pages} {121202(R)} (\bibinfo {year} {2010})}\BibitemShut {NoStop}%
\bibitem [{\citenamefont {Zapasskii}(2013)}]{VSZRev}%
  \BibitemOpen
  \bibfield  {author} {\bibinfo {author} {\bibfnamefont {V.~S.}\ \bibnamefont
  {Zapasskii}},\ }\href {\doibase 10.1364/AOP.5.000131} {\bibfield  {journal}
  {\bibinfo  {journal} {Adv. Opt. Photon.}\ }\textbf {\bibinfo {volume} {5}},\
  \bibinfo {pages} {131} (\bibinfo {year} {2013})}\BibitemShut {NoStop}%
\bibitem [{\citenamefont {Aleksandrov}\ and\ \citenamefont
  {Zapasskii}(1981)}]{Alexandrov81}%
  \BibitemOpen
  \bibfield  {author} {\bibinfo {author} {\bibfnamefont {E.~B.}\ \bibnamefont
  {Aleksandrov}}\ and\ \bibinfo {author} {\bibfnamefont {V.~S.}\ \bibnamefont
  {Zapasskii}},\ }\href@noop {} {\bibfield  {journal} {\bibinfo  {journal} {Zh.
  Eksp. Teor. Fiz. [Sov. Phys. JETP 54, 64 (1981)]}\ }\textbf {\bibinfo
  {volume} {81}},\ \bibinfo {pages} {132} (\bibinfo {year} {1981})}\BibitemShut
  {NoStop}%
\bibitem [{\citenamefont {Crooker}\ \emph {et~al.}(2004)\citenamefont
  {Crooker}, \citenamefont {Rickel}, \citenamefont {Balatsky},\ and\
  \citenamefont {Smith}}]{Crooker04}%
  \BibitemOpen
  \bibfield  {author} {\bibinfo {author} {\bibfnamefont {S.~A.}\ \bibnamefont
  {Crooker}}, \bibinfo {author} {\bibfnamefont {D.~G.}\ \bibnamefont {Rickel}},
  \bibinfo {author} {\bibfnamefont {A.~V.}\ \bibnamefont {Balatsky}}, \ and\
  \bibinfo {author} {\bibfnamefont {D.~L.}\ \bibnamefont {Smith}},\ }\href@noop
  {} {\bibfield  {journal} {\bibinfo  {journal} {Nature}\ }\textbf {\bibinfo
  {volume} {431}},\ \bibinfo {pages} {49} (\bibinfo {year} {2004})}\BibitemShut
  {NoStop}%
\bibitem [{\citenamefont {Oestreich}\ \emph {et~al.}(2005)\citenamefont
  {Oestreich}, \citenamefont {R\"omer}, \citenamefont {Haug},\ and\
  \citenamefont {H\"agele}}]{Oestreich05}%
  \BibitemOpen
  \bibfield  {author} {\bibinfo {author} {\bibfnamefont {M.}~\bibnamefont
  {Oestreich}}, \bibinfo {author} {\bibfnamefont {M.}~\bibnamefont {R\"omer}},
  \bibinfo {author} {\bibfnamefont {R.~J.}\ \bibnamefont {Haug}}, \ and\
  \bibinfo {author} {\bibfnamefont {D.}~\bibnamefont {H\"agele}},\ }\href@noop
  {} {\bibfield  {journal} {\bibinfo  {journal} {Phys. Rev. Lett.}\ }\textbf
  {\bibinfo {volume} {95}},\ \bibinfo {pages} {216603} (\bibinfo {year}
  {2005})}\BibitemShut {NoStop}%
\bibitem [{\citenamefont {M\"uller}\ \emph {et~al.}(2008)\citenamefont
  {M\"uller}, \citenamefont {R\"omer}, \citenamefont {Schuh}, \citenamefont
  {Wegscheider}, \citenamefont {H\"ubner},\ and\ \citenamefont
  {Oestreich}}]{Mueller08}%
  \BibitemOpen
  \bibfield  {author} {\bibinfo {author} {\bibfnamefont {G.~M.}\ \bibnamefont
  {M\"uller}}, \bibinfo {author} {\bibfnamefont {M.}~\bibnamefont {R\"omer}},
  \bibinfo {author} {\bibfnamefont {D.}~\bibnamefont {Schuh}}, \bibinfo
  {author} {\bibfnamefont {W.}~\bibnamefont {Wegscheider}}, \bibinfo {author}
  {\bibfnamefont {J.}~\bibnamefont {H\"ubner}}, \ and\ \bibinfo {author}
  {\bibfnamefont {M.}~\bibnamefont {Oestreich}},\ }\href {\doibase
  10.1103/PhysRevLett.101.206601} {\bibfield  {journal} {\bibinfo  {journal}
  {Phys. Rev. Lett.}\ }\textbf {\bibinfo {volume} {101}},\ \bibinfo {pages}
  {206601} (\bibinfo {year} {2008})}\BibitemShut {NoStop}%
\bibitem [{\citenamefont {Crooker}\ \emph {et~al.}(2009)\citenamefont
  {Crooker}, \citenamefont {Cheng},\ and\ \citenamefont {Smith}}]{Crooker09}%
  \BibitemOpen
  \bibfield  {author} {\bibinfo {author} {\bibfnamefont {S.~A.}\ \bibnamefont
  {Crooker}}, \bibinfo {author} {\bibfnamefont {L.}~\bibnamefont {Cheng}}, \
  and\ \bibinfo {author} {\bibfnamefont {D.~L.}\ \bibnamefont {Smith}},\ }\href
  {\doibase 10.1103/PhysRevB.79.035208} {\bibfield  {journal} {\bibinfo
  {journal} {Phys. Rev. B}\ }\textbf {\bibinfo {volume} {79}},\ \bibinfo
  {pages} {035208} (\bibinfo {year} {2009})}\BibitemShut {NoStop}%
\bibitem [{\citenamefont {Crooker}\ \emph {et~al.}(2010)\citenamefont
  {Crooker}, \citenamefont {Brandt}, \citenamefont {Sandfort}, \citenamefont
  {Greilich}, \citenamefont {Yakovlev}, \citenamefont {Reuter}, \citenamefont
  {Wieck},\ and\ \citenamefont {Bayer}}]{CrookerPRL2010}%
  \BibitemOpen
  \bibfield  {author} {\bibinfo {author} {\bibfnamefont {S.~A.}\ \bibnamefont
  {Crooker}}, \bibinfo {author} {\bibfnamefont {J.}~\bibnamefont {Brandt}},
  \bibinfo {author} {\bibfnamefont {C.}~\bibnamefont {Sandfort}}, \bibinfo
  {author} {\bibfnamefont {A.}~\bibnamefont {Greilich}}, \bibinfo {author}
  {\bibfnamefont {D.~R.}\ \bibnamefont {Yakovlev}}, \bibinfo {author}
  {\bibfnamefont {D.}~\bibnamefont {Reuter}}, \bibinfo {author} {\bibfnamefont
  {A.~D.}\ \bibnamefont {Wieck}}, \ and\ \bibinfo {author} {\bibfnamefont
  {M.}~\bibnamefont {Bayer}},\ }\href {\doibase 10.1103/PhysRevLett.104.036601}
  {\bibfield  {journal} {\bibinfo  {journal} {Phys. Rev. Lett.}\ }\textbf
  {\bibinfo {volume} {104}},\ \bibinfo {pages} {036601} (\bibinfo {year}
  {2010})}\BibitemShut {NoStop}%
\bibitem [{\citenamefont {Dahbashi}\ \emph {et~al.}(2014)\citenamefont
  {Dahbashi}, \citenamefont {H\"ubner}, \citenamefont {Berski}, \citenamefont
  {Pierz},\ and\ \citenamefont {Oestreich}}]{DahbashiPRL14}%
  \BibitemOpen
  \bibfield  {author} {\bibinfo {author} {\bibfnamefont {R.}~\bibnamefont
  {Dahbashi}}, \bibinfo {author} {\bibfnamefont {J.}~\bibnamefont {H\"ubner}},
  \bibinfo {author} {\bibfnamefont {F.}~\bibnamefont {Berski}}, \bibinfo
  {author} {\bibfnamefont {K.}~\bibnamefont {Pierz}}, \ and\ \bibinfo {author}
  {\bibfnamefont {M.}~\bibnamefont {Oestreich}},\ }\href {\doibase
  10.1103/PhysRevLett.112.156601} {\bibfield  {journal} {\bibinfo  {journal}
  {Phys. Rev. Lett.}\ }\textbf {\bibinfo {volume} {112}},\ \bibinfo {pages}
  {156601} (\bibinfo {year} {2014})}\BibitemShut {NoStop}%
\bibitem [{\citenamefont {Goryca}\ \emph {et~al.}(2019)\citenamefont {Goryca},
  \citenamefont {Wilson}, \citenamefont {Dey}, \citenamefont {Xu},\ and\
  \citenamefont {Crooker}}]{Goryca19}%
  \BibitemOpen
  \bibfield  {author} {\bibinfo {author} {\bibfnamefont {M.}~\bibnamefont
  {Goryca}}, \bibinfo {author} {\bibfnamefont {N.~P.}\ \bibnamefont {Wilson}},
  \bibinfo {author} {\bibfnamefont {P.}~\bibnamefont {Dey}}, \bibinfo {author}
  {\bibfnamefont {X.}~\bibnamefont {Xu}}, \ and\ \bibinfo {author}
  {\bibfnamefont {S.~A.}\ \bibnamefont {Crooker}},\ }\href
  {https://advances.sciencemag.org/content/5/3/eaau4899} {\bibfield  {journal}
  {\bibinfo  {journal} {Science Advances}\ }\textbf {\bibinfo {volume} {5}}
  (\bibinfo {year} {2019})}\BibitemShut {NoStop}%
\bibitem [{\citenamefont {Balk}\ \emph {et~al.}(2018)\citenamefont {Balk},
  \citenamefont {Li}, \citenamefont {Gilbert}, \citenamefont {Unguris},
  \citenamefont {Sinitsyn},\ and\ \citenamefont {Crooker}}]{CrookerPRX18}%
  \BibitemOpen
  \bibfield  {author} {\bibinfo {author} {\bibfnamefont {A.~L.}\ \bibnamefont
  {Balk}}, \bibinfo {author} {\bibfnamefont {F.}~\bibnamefont {Li}}, \bibinfo
  {author} {\bibfnamefont {I.}~\bibnamefont {Gilbert}}, \bibinfo {author}
  {\bibfnamefont {J.}~\bibnamefont {Unguris}}, \bibinfo {author} {\bibfnamefont
  {N.~A.}\ \bibnamefont {Sinitsyn}}, \ and\ \bibinfo {author} {\bibfnamefont
  {S.~A.}\ \bibnamefont {Crooker}},\ }\href {\doibase
  10.1103/PhysRevX.8.031078} {\bibfield  {journal} {\bibinfo  {journal} {Phys.
  Rev. X}\ }\textbf {\bibinfo {volume} {8}},\ \bibinfo {pages} {031078}
  (\bibinfo {year} {2018})}\BibitemShut {NoStop}%
\bibitem [{\citenamefont {Giri}\ \emph {et~al.}(2012)\citenamefont {Giri},
  \citenamefont {Cronenberger}, \citenamefont {Vladimirova}, \citenamefont
  {Scalbert}, \citenamefont {Kavokin}, \citenamefont {Glazov}, \citenamefont
  {Nawrocki}, \citenamefont {Lema\^{\i}tre},\ and\ \citenamefont
  {Bloch}}]{GiriPRB12}%
  \BibitemOpen
  \bibfield  {author} {\bibinfo {author} {\bibfnamefont {R.}~\bibnamefont
  {Giri}}, \bibinfo {author} {\bibfnamefont {S.}~\bibnamefont {Cronenberger}},
  \bibinfo {author} {\bibfnamefont {M.}~\bibnamefont {Vladimirova}}, \bibinfo
  {author} {\bibfnamefont {D.}~\bibnamefont {Scalbert}}, \bibinfo {author}
  {\bibfnamefont {K.~V.}\ \bibnamefont {Kavokin}}, \bibinfo {author}
  {\bibfnamefont {M.~M.}\ \bibnamefont {Glazov}}, \bibinfo {author}
  {\bibfnamefont {M.}~\bibnamefont {Nawrocki}}, \bibinfo {author}
  {\bibfnamefont {A.}~\bibnamefont {Lema\^{\i}tre}}, \ and\ \bibinfo {author}
  {\bibfnamefont {J.}~\bibnamefont {Bloch}},\ }\href {\doibase
  10.1103/PhysRevB.85.195313} {\bibfield  {journal} {\bibinfo  {journal} {Phys.
  Rev. B}\ }\textbf {\bibinfo {volume} {85}},\ \bibinfo {pages} {195313}
  (\bibinfo {year} {2012})}\BibitemShut {NoStop}%
\bibitem [{\citenamefont {Gorbovitskii}\ and\ \citenamefont
  {Perel}(1983)}]{Gorbovitskii83}%
  \BibitemOpen
  \bibfield  {author} {\bibinfo {author} {\bibfnamefont {B.}~\bibnamefont
  {Gorbovitskii}}\ and\ \bibinfo {author} {\bibfnamefont {V.}~\bibnamefont
  {Perel}},\ }\href@noop {} {\bibfield  {journal} {\bibinfo  {journal} {Opt.
  Spectrosc.}\ }\textbf {\bibinfo {volume} {54}},\ \bibinfo {pages} {388}
  (\bibinfo {year} {1983})}\BibitemShut {NoStop}%
\bibitem [{\citenamefont {Glazov}\ and\ \citenamefont
  {Zapasskii}(2015)}]{GlazovOE15}%
  \BibitemOpen
  \bibfield  {author} {\bibinfo {author} {\bibfnamefont {M.~M.}\ \bibnamefont
  {Glazov}}\ and\ \bibinfo {author} {\bibfnamefont {V.~S.}\ \bibnamefont
  {Zapasskii}},\ }\href {\doibase 10.1364/OE.23.011713} {\bibfield  {journal}
  {\bibinfo  {journal} {Opt. Express}\ }\textbf {\bibinfo {volume} {23}},\
  \bibinfo {pages} {11713} (\bibinfo {year} {2015})}\BibitemShut {NoStop}%
\bibitem [{\citenamefont {Scalbert}(2019)}]{ScalbertPRB19}%
  \BibitemOpen
  \bibfield  {author} {\bibinfo {author} {\bibfnamefont {D.}~\bibnamefont
  {Scalbert}},\ }\href {\doibase 10.1103/PhysRevB.99.205305} {\bibfield
  {journal} {\bibinfo  {journal} {Phys. Rev. B}\ }\textbf {\bibinfo {volume}
  {99}},\ \bibinfo {pages} {205305} (\bibinfo {year} {2019})}\BibitemShut
  {NoStop}%
\bibitem [{\citenamefont {Cronenberger}\ \emph {et~al.}(2019)\citenamefont
  {Cronenberger}, \citenamefont {Abbas}, \citenamefont {Scalbert},\ and\
  \citenamefont {Boukari}}]{STSNSPRL19}%
  \BibitemOpen
  \bibfield  {author} {\bibinfo {author} {\bibfnamefont {S.}~\bibnamefont
  {Cronenberger}}, \bibinfo {author} {\bibfnamefont {C.}~\bibnamefont {Abbas}},
  \bibinfo {author} {\bibfnamefont {D.}~\bibnamefont {Scalbert}}, \ and\
  \bibinfo {author} {\bibfnamefont {H.}~\bibnamefont {Boukari}},\ }\href
  {\doibase 10.1103/PhysRevLett.123.017401} {\bibfield  {journal} {\bibinfo
  {journal} {Phys. Rev. Lett.}\ }\textbf {\bibinfo {volume} {123}},\ \bibinfo
  {pages} {017401} (\bibinfo {year} {2019})}\BibitemShut {NoStop}%
\bibitem [{\citenamefont {Dahbashi}\ \emph {et~al.}(2012)\citenamefont
  {Dahbashi}, \citenamefont {Hübner}, \citenamefont {Berski}, \citenamefont
  {Wiegand}, \citenamefont {Marie}, \citenamefont {Pierz}, \citenamefont
  {Schumacher},\ and\ \citenamefont {Oestreich}}]{DahbashiAPL12}%
  \BibitemOpen
  \bibfield  {author} {\bibinfo {author} {\bibfnamefont {R.}~\bibnamefont
  {Dahbashi}}, \bibinfo {author} {\bibfnamefont {J.}~\bibnamefont {Hübner}},
  \bibinfo {author} {\bibfnamefont {F.}~\bibnamefont {Berski}}, \bibinfo
  {author} {\bibfnamefont {J.}~\bibnamefont {Wiegand}}, \bibinfo {author}
  {\bibfnamefont {X.}~\bibnamefont {Marie}}, \bibinfo {author} {\bibfnamefont
  {K.}~\bibnamefont {Pierz}}, \bibinfo {author} {\bibfnamefont {H.~W.}\
  \bibnamefont {Schumacher}}, \ and\ \bibinfo {author} {\bibfnamefont
  {M.}~\bibnamefont {Oestreich}},\ }\href@noop {} {\bibfield  {journal}
  {\bibinfo  {journal} {App. Phys. Lett.}\ }\textbf {\bibinfo {volume} {100}},\
  \bibinfo {pages} {031906} (\bibinfo {year} {2012})}\BibitemShut {NoStop}%
\bibitem [{\citenamefont {Glasenapp}\ \emph {et~al.}(2013)\citenamefont
  {Glasenapp}, \citenamefont {Greilich}, \citenamefont {Ryzhov}, \citenamefont
  {Zapasskii}, \citenamefont {Yakovlev}, \citenamefont {Kozlov},\ and\
  \citenamefont {Bayer}}]{GlasenappPRB13}%
  \BibitemOpen
  \bibfield  {author} {\bibinfo {author} {\bibfnamefont {P.}~\bibnamefont
  {Glasenapp}}, \bibinfo {author} {\bibfnamefont {A.}~\bibnamefont {Greilich}},
  \bibinfo {author} {\bibfnamefont {I.~I.}\ \bibnamefont {Ryzhov}}, \bibinfo
  {author} {\bibfnamefont {V.~S.}\ \bibnamefont {Zapasskii}}, \bibinfo {author}
  {\bibfnamefont {D.~R.}\ \bibnamefont {Yakovlev}}, \bibinfo {author}
  {\bibfnamefont {G.~G.}\ \bibnamefont {Kozlov}}, \ and\ \bibinfo {author}
  {\bibfnamefont {M.}~\bibnamefont {Bayer}},\ }\href {\doibase
  10.1103/PhysRevB.88.165314} {\bibfield  {journal} {\bibinfo  {journal} {Phys.
  Rev. B}\ }\textbf {\bibinfo {volume} {88}},\ \bibinfo {pages} {165314}
  (\bibinfo {year} {2013})}\BibitemShut {NoStop}%
\bibitem [{\citenamefont {Zapasskii}\ and\ \citenamefont
  {Przhibelskii}(2011)}]{Zapasskii2011}%
  \BibitemOpen
  \bibfield  {author} {\bibinfo {author} {\bibfnamefont {V.~S.}\ \bibnamefont
  {Zapasskii}}\ and\ \bibinfo {author} {\bibfnamefont {S.~G.}\ \bibnamefont
  {Przhibelskii}},\ }\href@noop {} {\bibfield  {journal} {\bibinfo  {journal}
  {Opt. Spectrosc.}\ }\textbf {\bibinfo {volume} {110}},\ \bibinfo {pages}
  {917} (\bibinfo {year} {2011})}\BibitemShut {NoStop}%
\bibitem [{\citenamefont {Poltavtsev}\ \emph {et~al.}(2014)\citenamefont
  {Poltavtsev}, \citenamefont {Ryzhov}, \citenamefont {Glazov}, \citenamefont
  {Kozlov}, \citenamefont {Zapasskii}, \citenamefont {Kavokin}, \citenamefont
  {Lagoudakis}, \citenamefont {Smirnov},\ and\ \citenamefont
  {Ivchenko}}]{PoltavtsevPRB14}%
  \BibitemOpen
  \bibfield  {author} {\bibinfo {author} {\bibfnamefont {S.~V.}\ \bibnamefont
  {Poltavtsev}}, \bibinfo {author} {\bibfnamefont {I.~I.}\ \bibnamefont
  {Ryzhov}}, \bibinfo {author} {\bibfnamefont {M.~M.}\ \bibnamefont {Glazov}},
  \bibinfo {author} {\bibfnamefont {G.~G.}\ \bibnamefont {Kozlov}}, \bibinfo
  {author} {\bibfnamefont {V.~S.}\ \bibnamefont {Zapasskii}}, \bibinfo {author}
  {\bibfnamefont {A.~V.}\ \bibnamefont {Kavokin}}, \bibinfo {author}
  {\bibfnamefont {P.~G.}\ \bibnamefont {Lagoudakis}}, \bibinfo {author}
  {\bibfnamefont {D.~S.}\ \bibnamefont {Smirnov}}, \ and\ \bibinfo {author}
  {\bibfnamefont {E.~L.}\ \bibnamefont {Ivchenko}},\ }\href {\doibase
  10.1103/PhysRevB.89.081304} {\bibfield  {journal} {\bibinfo  {journal} {Phys.
  Rev. B}\ }\textbf {\bibinfo {volume} {89}},\ \bibinfo {pages} {081304(R)}
  (\bibinfo {year} {2014})}\BibitemShut {NoStop}%
\bibitem [{\citenamefont {Cronenberger}\ and\ \citenamefont
  {Scalbert}(2016)}]{Cronenberger16}%
  \BibitemOpen
  \bibfield  {author} {\bibinfo {author} {\bibfnamefont {S.}~\bibnamefont
  {Cronenberger}}\ and\ \bibinfo {author} {\bibfnamefont {D.}~\bibnamefont
  {Scalbert}},\ }\href {\doibase 10.1063/1.4962863} {\bibfield  {journal}
  {\bibinfo  {journal} {Rev. Sci. Instrum.}\ }\textbf {\bibinfo {volume}
  {87}},\ \bibinfo {pages} {093111} (\bibinfo {year} {2016})}\BibitemShut
  {NoStop}%
\bibitem [{\citenamefont {Sterin}\ \emph {et~al.}(2018)\citenamefont {Sterin},
  \citenamefont {Wiegand}, \citenamefont {H\"ubner},\ and\ \citenamefont
  {Oestreich}}]{Sterin18}%
  \BibitemOpen
  \bibfield  {author} {\bibinfo {author} {\bibfnamefont {P.}~\bibnamefont
  {Sterin}}, \bibinfo {author} {\bibfnamefont {J.}~\bibnamefont {Wiegand}},
  \bibinfo {author} {\bibfnamefont {J.}~\bibnamefont {H\"ubner}}, \ and\
  \bibinfo {author} {\bibfnamefont {M.}~\bibnamefont {Oestreich}},\ }\href
  {\doibase 10.1103/PhysRevApplied.9.034003} {\bibfield  {journal} {\bibinfo
  {journal} {Phys. Rev. Applied}\ }\textbf {\bibinfo {volume} {9}},\ \bibinfo
  {pages} {034003} (\bibinfo {year} {2018})}\BibitemShut {NoStop}%
\bibitem [{\citenamefont {Petrov}\ \emph {et~al.}(2018)\citenamefont {Petrov},
  \citenamefont {Kamenskii}, \citenamefont {Zapasskii}, \citenamefont {Bayer},\
  and\ \citenamefont {Greilich}}]{Homodyne2018}%
  \BibitemOpen
  \bibfield  {author} {\bibinfo {author} {\bibfnamefont {M.~Y.}\ \bibnamefont
  {Petrov}}, \bibinfo {author} {\bibfnamefont {A.~N.}\ \bibnamefont
  {Kamenskii}}, \bibinfo {author} {\bibfnamefont {V.~S.}\ \bibnamefont
  {Zapasskii}}, \bibinfo {author} {\bibfnamefont {M.}~\bibnamefont {Bayer}}, \
  and\ \bibinfo {author} {\bibfnamefont {A.}~\bibnamefont {Greilich}},\ }\href
  {\doibase 10.1103/PhysRevB.97.125202} {\bibfield  {journal} {\bibinfo
  {journal} {Phys. Rev. B}\ }\textbf {\bibinfo {volume} {97}},\ \bibinfo
  {pages} {125202} (\bibinfo {year} {2018})}\BibitemShut {NoStop}%
\bibitem [{\citenamefont {Kozlov}\ \emph {et~al.}(2017)\citenamefont {Kozlov},
  \citenamefont {Ryzhov},\ and\ \citenamefont {Zapasskii}}]{KozlovPRA17}%
  \BibitemOpen
  \bibfield  {author} {\bibinfo {author} {\bibfnamefont {G.~G.}\ \bibnamefont
  {Kozlov}}, \bibinfo {author} {\bibfnamefont {I.~I.}\ \bibnamefont {Ryzhov}},
  \ and\ \bibinfo {author} {\bibfnamefont {V.~S.}\ \bibnamefont {Zapasskii}},\
  }\href {\doibase 10.1103/PhysRevA.95.043810} {\bibfield  {journal} {\bibinfo
  {journal} {Phys. Rev. A}\ }\textbf {\bibinfo {volume} {95}},\ \bibinfo
  {pages} {043810} (\bibinfo {year} {2017})}\BibitemShut {NoStop}%
\bibitem [{\citenamefont {Ryzhov}\ \emph {et~al.}(2015)\citenamefont {Ryzhov},
  \citenamefont {Poltavtsev}, \citenamefont {Kozlov}, \citenamefont {Kavokin},
  \citenamefont {Lagoudakis},\ and\ \citenamefont {Zapasskii}}]{RyzhovJAP15}%
  \BibitemOpen
  \bibfield  {author} {\bibinfo {author} {\bibfnamefont {I.~I.}\ \bibnamefont
  {Ryzhov}}, \bibinfo {author} {\bibfnamefont {S.~V.}\ \bibnamefont
  {Poltavtsev}}, \bibinfo {author} {\bibfnamefont {G.~G.}\ \bibnamefont
  {Kozlov}}, \bibinfo {author} {\bibfnamefont {A.~V.}\ \bibnamefont {Kavokin}},
  \bibinfo {author} {\bibfnamefont {P.~V.}\ \bibnamefont {Lagoudakis}}, \ and\
  \bibinfo {author} {\bibfnamefont {V.~S.}\ \bibnamefont {Zapasskii}},\ }\href
  {\doibase 10.1063/1.4922405} {\bibfield  {journal} {\bibinfo  {journal} {J.
  Appl. Phys.}\ }\textbf {\bibinfo {volume} {117}},\ \bibinfo {pages} {224305}
  (\bibinfo {year} {2015})},\ \Eprint
  {http://arxiv.org/abs/https://doi.org/10.1063/1.4922405}
  {https://doi.org/10.1063/1.4922405} \BibitemShut {NoStop}%
\bibitem [{\citenamefont {Scalbert}(2016)}]{Scalbert2016}%
  \BibitemOpen
  \bibfield  {author} {\bibinfo {author} {\bibfnamefont {D.}~\bibnamefont
  {Scalbert}},\ }\href@noop {} {\bibfield  {journal} {\bibinfo  {journal}
  {Superlattice Microst.}\ }\textbf {\bibinfo {volume} {92}},\ \bibinfo {pages}
  {348} (\bibinfo {year} {2016})}\BibitemShut {NoStop}%
\bibitem [{\citenamefont {Glasenapp}\ \emph {et~al.}(2016)\citenamefont
  {Glasenapp}, \citenamefont {Smirnov}, \citenamefont {Greilich}, \citenamefont
  {Hackmann}, \citenamefont {Glazov}, \citenamefont {Anders},\ and\
  \citenamefont {Bayer}}]{GlasenappPRB16}%
  \BibitemOpen
  \bibfield  {author} {\bibinfo {author} {\bibfnamefont {P.}~\bibnamefont
  {Glasenapp}}, \bibinfo {author} {\bibfnamefont {D.~S.}\ \bibnamefont
  {Smirnov}}, \bibinfo {author} {\bibfnamefont {A.}~\bibnamefont {Greilich}},
  \bibinfo {author} {\bibfnamefont {J.}~\bibnamefont {Hackmann}}, \bibinfo
  {author} {\bibfnamefont {M.~M.}\ \bibnamefont {Glazov}}, \bibinfo {author}
  {\bibfnamefont {F.~B.}\ \bibnamefont {Anders}}, \ and\ \bibinfo {author}
  {\bibfnamefont {M.}~\bibnamefont {Bayer}},\ }\href {\doibase
  10.1103/PhysRevB.93.205429} {\bibfield  {journal} {\bibinfo  {journal} {Phys.
  Rev. B}\ }\textbf {\bibinfo {volume} {93}},\ \bibinfo {pages} {205429}
  (\bibinfo {year} {2016})}\BibitemShut {NoStop}%
\bibitem [{\citenamefont {Dzhioev}\ \emph {et~al.}(2002)\citenamefont
  {Dzhioev}, \citenamefont {Kavokin}, \citenamefont {Korenev}, \citenamefont
  {Lazarev}, \citenamefont {Meltser}, \citenamefont {Stepanova}, \citenamefont
  {Zakharchenya}, \citenamefont {Gammon},\ and\ \citenamefont
  {Katzer}}]{DzhioevPRB02}%
  \BibitemOpen
  \bibfield  {author} {\bibinfo {author} {\bibfnamefont {R.~I.}\ \bibnamefont
  {Dzhioev}}, \bibinfo {author} {\bibfnamefont {K.~V.}\ \bibnamefont
  {Kavokin}}, \bibinfo {author} {\bibfnamefont {V.~L.}\ \bibnamefont
  {Korenev}}, \bibinfo {author} {\bibfnamefont {M.~V.}\ \bibnamefont
  {Lazarev}}, \bibinfo {author} {\bibfnamefont {B.~Y.}\ \bibnamefont
  {Meltser}}, \bibinfo {author} {\bibfnamefont {M.~N.}\ \bibnamefont
  {Stepanova}}, \bibinfo {author} {\bibfnamefont {B.~P.}\ \bibnamefont
  {Zakharchenya}}, \bibinfo {author} {\bibfnamefont {D.}~\bibnamefont
  {Gammon}}, \ and\ \bibinfo {author} {\bibfnamefont {D.~S.}\ \bibnamefont
  {Katzer}},\ }\href {\doibase 10.1103/PhysRevB.66.245204} {\bibfield
  {journal} {\bibinfo  {journal} {Phys. Rev. B}\ }\textbf {\bibinfo {volume}
  {66}},\ \bibinfo {pages} {245204} (\bibinfo {year} {2002})}\BibitemShut
  {NoStop}%
\bibitem [{\citenamefont {Smirnov}\ \emph {et~al.}(2018)\citenamefont
  {Smirnov}, \citenamefont {Zhukov}, \citenamefont {Kirstein}, \citenamefont
  {Yakovlev}, \citenamefont {Reuter}, \citenamefont {Wieck}, \citenamefont
  {Bayer}, \citenamefont {Greilich},\ and\ \citenamefont
  {Glazov}}]{SmirnovPRB18}%
  \BibitemOpen
  \bibfield  {author} {\bibinfo {author} {\bibfnamefont {D.~S.}\ \bibnamefont
  {Smirnov}}, \bibinfo {author} {\bibfnamefont {E.~A.}\ \bibnamefont {Zhukov}},
  \bibinfo {author} {\bibfnamefont {E.}~\bibnamefont {Kirstein}}, \bibinfo
  {author} {\bibfnamefont {D.~R.}\ \bibnamefont {Yakovlev}}, \bibinfo {author}
  {\bibfnamefont {D.}~\bibnamefont {Reuter}}, \bibinfo {author} {\bibfnamefont
  {A.~D.}\ \bibnamefont {Wieck}}, \bibinfo {author} {\bibfnamefont
  {M.}~\bibnamefont {Bayer}}, \bibinfo {author} {\bibfnamefont
  {A.}~\bibnamefont {Greilich}}, \ and\ \bibinfo {author} {\bibfnamefont
  {M.~M.}\ \bibnamefont {Glazov}},\ }\href {\doibase
  10.1103/PhysRevB.98.125306} {\bibfield  {journal} {\bibinfo  {journal} {Phys.
  Rev. B}\ }\textbf {\bibinfo {volume} {98}},\ \bibinfo {pages} {125306}
  (\bibinfo {year} {2018})}\BibitemShut {NoStop}%
\bibitem [{\citenamefont {Li}\ \emph {et~al.}(2012)\citenamefont {Li},
  \citenamefont {Sinitsyn}, \citenamefont {Smith}, \citenamefont {Reuter},
  \citenamefont {Wieck}, \citenamefont {Yakovlev}, \citenamefont {Bayer},\ and\
  \citenamefont {Crooker}}]{LiPRL12}%
  \BibitemOpen
  \bibfield  {author} {\bibinfo {author} {\bibfnamefont {Y.}~\bibnamefont
  {Li}}, \bibinfo {author} {\bibfnamefont {N.}~\bibnamefont {Sinitsyn}},
  \bibinfo {author} {\bibfnamefont {D.~L.}\ \bibnamefont {Smith}}, \bibinfo
  {author} {\bibfnamefont {D.}~\bibnamefont {Reuter}}, \bibinfo {author}
  {\bibfnamefont {A.~D.}\ \bibnamefont {Wieck}}, \bibinfo {author}
  {\bibfnamefont {D.~R.}\ \bibnamefont {Yakovlev}}, \bibinfo {author}
  {\bibfnamefont {M.}~\bibnamefont {Bayer}}, \ and\ \bibinfo {author}
  {\bibfnamefont {S.~A.}\ \bibnamefont {Crooker}},\ }\href {\doibase
  10.1103/PhysRevLett.108.186603} {\bibfield  {journal} {\bibinfo  {journal}
  {Phys. Rev. Lett.}\ }\textbf {\bibinfo {volume} {108}},\ \bibinfo {pages}
  {186603} (\bibinfo {year} {2012})}\BibitemShut {NoStop}%
\bibitem [{\citenamefont {Karrai}\ and\ \citenamefont
  {Warburton}(2003)}]{KarraiSM03}%
  \BibitemOpen
  \bibfield  {author} {\bibinfo {author} {\bibfnamefont {K.}~\bibnamefont
  {Karrai}}\ and\ \bibinfo {author} {\bibfnamefont {R.~J.}\ \bibnamefont
  {Warburton}},\ }\href@noop {} {\bibfield  {journal} {\bibinfo  {journal}
  {Superlattice Microst.}\ }\textbf {\bibinfo {volume} {33}},\ \bibinfo {pages}
  {311 } (\bibinfo {year} {2003})}\BibitemShut {NoStop}%
\bibitem [{\citenamefont {Nguyen}\ \emph {et~al.}(2011)\citenamefont {Nguyen},
  \citenamefont {Sallen}, \citenamefont {Voisin}, \citenamefont {Roussignol},
  \citenamefont {Diederichs},\ and\ \citenamefont {Cassabois}}]{NguyenAPL11}%
  \BibitemOpen
  \bibfield  {author} {\bibinfo {author} {\bibfnamefont {H.~S.}\ \bibnamefont
  {Nguyen}}, \bibinfo {author} {\bibfnamefont {G.}~\bibnamefont {Sallen}},
  \bibinfo {author} {\bibfnamefont {C.}~\bibnamefont {Voisin}}, \bibinfo
  {author} {\bibfnamefont {P.}~\bibnamefont {Roussignol}}, \bibinfo {author}
  {\bibfnamefont {C.}~\bibnamefont {Diederichs}}, \ and\ \bibinfo {author}
  {\bibfnamefont {G.}~\bibnamefont {Cassabois}},\ }\href@noop {} {\bibfield
  {journal} {\bibinfo  {journal} {Appl. Phys. Lett.}\ }\textbf {\bibinfo
  {volume} {99}},\ \bibinfo {pages} {261904} (\bibinfo {year}
  {2011})}\BibitemShut {NoStop}%
\bibitem [{\citenamefont {Matthiesen}\ \emph {et~al.}(2012)\citenamefont
  {Matthiesen}, \citenamefont {Vamivakas},\ and\ \citenamefont
  {Atat\"ure}}]{MatthiesenPRL12}%
  \BibitemOpen
  \bibfield  {author} {\bibinfo {author} {\bibfnamefont {C.}~\bibnamefont
  {Matthiesen}}, \bibinfo {author} {\bibfnamefont {A.~N.}\ \bibnamefont
  {Vamivakas}}, \ and\ \bibinfo {author} {\bibfnamefont {M.}~\bibnamefont
  {Atat\"ure}},\ }\href {\doibase 10.1103/PhysRevLett.108.093602} {\bibfield
  {journal} {\bibinfo  {journal} {Phys. Rev. Lett.}\ }\textbf {\bibinfo
  {volume} {108}},\ \bibinfo {pages} {093602} (\bibinfo {year}
  {2012})}\BibitemShut {NoStop}%
\end{thebibliography}

\providecommand{\noopsort}[1]{}\providecommand{\singleletter}[1]{#1}%

\end{document}